\documentclass[%
 preprint,
groupedaddress,
superscriptaddress,
 amsmath,amssymb,
 aps,
 prx,
 showpacs
]{revtex4-2}

\usepackage{graphicx}
\usepackage{makecell,booktabs}
\usepackage{dcolumn}
\usepackage{bm}
\usepackage{hyperref}
\hypersetup{
    colorlinks=true,
    urlcolor= blue,
    citecolor=blue,
linkcolor= blue}
\usepackage{xcolor}
\usepackage{comment}

\begin{document}


\title{Local ferromagnetic resonance measurements of mesoscopically patterned ferromagnets using deterministically placed nanodiamonds }

\author{Jeffrey Rable}
\author{Benjamin Piazza}
\altaffiliation{ 
Current address: Network Science Institute, Northeastern University, Boston, MA 02115, USA
}
\author{Jyotirmay Dwivedi}
\author{Nitin Samarth}
\email{nsamarth@psu.edu}
\affiliation{Department of Physics, Pennsylvania State University, University Park, Pennsylvania 16802, USA}

\date{\today}

\begin{abstract}
Nitrogen-vacancy centers in diamond have recently been established as effective sensors of the magnetization dynamics in vicinal ferromagnetic materials. We demonstrate sub-100 nm placement accuracy of nitrogen-vacancy-containing nanodiamonds and use these as local sensors that probe optically detected ferromagnetic resonance in mesoscopically patterned Permalloy islands. These measurements reveal variations in the ferromagnetic resonance signal at different sites on these structures with distinct behavior in the edge and the bulk of patterned features. These test measurements establish an easily implemented approach for spatially targeted measurements of spin dynamics in mesoscale ferromagnets. In principle, the methodology can also be extended to local studies of nanoscale ferromagnets such as single magnetic nanowires and nanoparticles.  
\end{abstract}

\maketitle

\section{\label{sec:level1}Introduction}
Contemporary problems of interest in spintronics often require knowledge of the dynamical behavior of magnonic (spin wave) excitations in patterned ferromagnetic devices. For example, this information is important in the creation and characterization of magnon quantum buses \cite{Candido2020,Fukami2021} and other magnonic devices such as spin-based transistors \cite{Chumak2014}. The nitrogen-vacancy (NV) center in diamond has emerged as an effective non-perturbative local probe for characterizing the magnetic properties of such systems \cite{Casola_2018_NatRevMat}. NV center-based local magnetometry has provided new insights into static spin configurations of skyrmions \cite{Dovzhenko2018,Jenkins2019} and magnetic domain walls \cite{Tetienne2015}, as well as the dynamical behavior of magnons \cite{Wolfe2014, Du2017, Andrich2017,Lee-Wong2020} and vortices \cite{trimble2021}. In the latter context, it is important to develop techniques that allow local measurements of ferromagnetic resonance (FMR) at targeted locations in a ferromagnetic sample or device. 

Continuous wave NV center-based optically detected ferromagnetic resonance (ODFMR) measurements rely on the quenching of the NV center fluorescence when a vicinal ferromagnet meets the conditions for FMR \cite{Wolfe2014}. This is attributed to an increased magnon density and an accompanying enhancement of the magnetic field noise sensed by the NV centers, thus leading to increased spin relaxation  \cite{Du2017}. Local measurements of ODFMR have relied on stochastic distribution of dropcast nanodiamonds \cite{Wolfe2014,Page2019}, stochastic distribution of NVs in a diamond film \cite{VanderSar2015,Page2019,trimble2021,Purser2020}, proximate placement of a diamond nanobeam \cite{Du2017,Lee-Wong2020}, and chemically patterned directed assembly of nanodiamonds \cite{Andrich2017}. In principle, scanned probe NV center magnetometry could provide a means of carrying out local ODFMR with imaging capability. However, since NV detection of ODFMR rapidly decreases in sensitivity with increasing sample-probe distance \cite{Purser2020}, NV scanning probe measurements of ODFMR may be constrained by the tip-sample distances that are typically greater than about 100 nm \cite{Tetienne2015,Rohner2019,Yu2018}. Although one scanning NV technique allows for smaller tip-sample separation ($\sim 30$ nm) \cite{Dovzhenko2018}, it would be technically challenging to engineer effective excitation of the FMR in a ferromagnetic sample in this geometry. This may account for the absence (as yet) of any published reports of ODFMR using a scanning NV center probe. As an aside, we note that local magnetization dynamics of ferromagnets can also be effectively probed and imaged via a completely different method, namely scanning ferromagnetic resonance force microscopy (FMRFM) which uses a microscale force cantilever to detect FMR with $\sim100 - 200$ nm spatial resolution \cite{Lee_Nature_2010,Yu_WOS:000255962500061,Guo_PhysRevLett.110.017601,Wu_PhysRevB.101.184409}. As with NV center based ODFMR, the FMRFM technique is amenable to measurements at ambient temperature. However, in contrast with NV center ODMFR, the FMRFM method requires more sophisticated instrumentation and it is more cumbersome to collect FMR data that densely spans frequency-magnetic field space. Thus, a question of interest is whether one can develop a simpler approach for locally measuring FMR at targeted sites on a mesoscopic or nanoscale ferromagnetic structure without having to resort to the sophisticated instrumentation required by scanning microscopy techniques.

In this paper, we demonstrate the use of an atomic force microscope (AFM) to achieve well-controlled positioning of NV-containing nanodiamonds as FMR sensors,  with sizes between 40 nm - 100 nm and with sub-100 nm accuracy. We use these deterministically placed nanodiamonds to perform local ODFMR measurements of mesoscale ($5 - 10 \mu$m lateral size) features patterned in ferromagnetic (Permalloy, Py) thin films. Although similar to a 'pick-and-place' technique previously reported for assembling nanodiamonds at desired locations \cite{Schell2011,Bogdanov2019}, this approach has not yet been exploited to probe the localized magnetization dynamics of magnetic materials. We note that chemically patterned directed assembly of nanodiamonds has been effectively used for probing ODFMR in YIG devices \cite{Andrich2017}; however, the measured locations are constrained in advance by lithographic patterning and subject to overlay error. We seek a more flexible approach that allows the measured locations to be varied at will. The placement precision in our proof-of-concept demonstration can, in principle, also allow for the targeted measurement of smaller nanoscale structures, such as nanowires, or of localized modes in larger structures, such as edge modes and defect modes.    

\section{\label{sec:level2}Methods}

\subsection{\label{sec:level3}Nanodiamond Placement and ODFMR}

\begin{figure}
\includegraphics{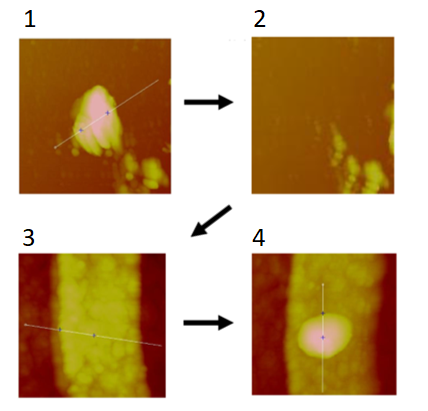}
\caption{\label{fig:NDPlacement} Demonstration of the nanodiamond placement process with a 40 nm diameter nanodiamond (1). After lifting the nanodiamond and confirming it is no longer on the sample (2), we move over to the deposition site, a 100 nm wide nanowire (3). Then, we can deposit via ramp or lift mode and confirm placement via another scan (4).} 
\end{figure}

We first describe the experimental approach for deterministic placement. We begin with a sample containing patterned permalloy structures in the center and drop cast an aqueous solution of 100 nm diameter nanodiamonds with 3 ppm NV centers onto the edge of the sample. This is done to avoid directly depositing nanodiamonds onto the features. Next, we scan the area where the nanodiamonds were drop cast using a Veeco Nanoscope IIIA Multimode AFM with a gold coated silicon tip, which increases the probability of pickup over a standard silicon tip, likely because of stronger van der Waals forces or malleability of the gold. When we find a particle that matches the dimensions of the dispersed nanodiamonds, we zoom in on the particle (Fig. \ref{fig:NDPlacement}, step 1) and enter ramp mode. Then, we ramp into the nanodiamond and re-scan the area where it was in the prior scan to confirm pickup (Fig. \ref{fig:NDPlacement}, step 2). If the nanodiamond remains at the site, we ramp in again until it is picked up or attempt pickup on a different particle.

If the nanodiamond does not appear on the post-ramp scan, indicating that it was picked up, we can begin the placement process by moving the AFM tip to our patterned features. With the nanodiamond still attached to the tip, we can scan the sample looking for the feature that we want to deposit the particle on (Fig. \ref{fig:NDPlacement}, step 3). When we find the device, we either repeat the process used in pickup, ramping into the feature until the nanodiamond dislodges, or we scan across the sample at a constant height below the surface using the AFM's lift mode, scraping the nanodiamond-coated tip along until it dislodges. While this second method works more consistently, it risks scratching the sample if the tip is dug into the feature. Finally, with the diamond dislodged, we perform a final scan to confirm that the nanodiamond is in the desired location (Fig. \ref{fig:NDPlacement}, step 4). We caution that we do not yet have a systematic measure of the success rate of the method. The repeatability of the technique appears to depend on factors that are not completely understood and varies with the details of the drop casting and the substrate.   

For optical polarization and readout of the NV center fluorescence, we used a 1 mW, 532 nm continuous wave laser and an ID Quantique ID100 avalanche photodiode. A scanning mirror scans across the surface for imaging and allows focusing on a site for ODFMR measurements. A static magnetic field is applied using a permanent N52 magnet mounted on a highly repeatable stepper motor linear stage. The applied field is calibrated in the plane of the sample using a single crystal diamond film containing NV centers; this is achieved using the known orientations and the Zeeman splitting of the NV electronic ground state spin transition. During the ODFMR measurements, a microwave magnetic field is applied via a 25 \textmu m diameter gold wire run across the sample. This microwave field both drives FMR in the magnetic features and the NV spin state transitions. Additionally, when FMR is driven in the sample, new, higher frequency magnons are generated via scattering and thermal mechanisms. The dipolar field noise generated by these incoherent magnons also affects the NV spin state transitions; this effect is believed to be responsible for the detection of FMR via an optical contrast even though the FMR is driven at frequencies off-resonant from those that drive the NV spin transitions \cite{Du2017}.


\subsection{\label{sec:level4}Micromagnetic Simulations}

Prior to our measurements, we performed micromagnetic simulations of the FMR modes in patterned Py features identical to those measured experimentally. The Py film thickness in all these simulations is 10 nm. The simulations were performed with the Mumax3 software package, which uses the Landau-Lifshitz-Gilbert equation:
	
\begin{equation*}
    \frac{\partial \vec{M}}{\partial t} = \gamma_{\text{LL}} \frac{1}{1+\alpha^2} (\vec{m} \times \vec{B_{\text{eff}}} + \alpha(\vec{m} \times (\vec{m} \times \vec{B_{\text{eff}}}))
    \tag{1}
    \label{eqn:LLG}
\end{equation*}

 \noindent to calculate the evolution of the magnetization $\vec{M}$ of finite ferromagnetic cells. In Eq. \ref{eqn:LLG}, $\alpha$ is the Gilbert damping of the material, $\gamma_{\text{LL}}$ is the gyromagnetic ratio of the material, and $B_{eff}$ is the effective magnetic field at that cell, which includes contributions from external, demagnetization, exchange, and anisotropy fields \cite{Vansteenkiste2014}.

The simulations were performed using 5 nm x 5 nm x 10 nm cells and the geometries consisted of permalloy features with the parameters in table \ref{tab:PyParameters}.

\begin{table}[ht]
\centering
\begin{tabular}{|l|l|l|}
\hline
Parameter & Value \\
\hline
$M_{\text{s}}$ & 8  x $10^{5}$ A/m\\
\hline
$A_{\text{ex}}$ & 1.3 x $10^{-11}$ J/m\\
\hline
$\alpha$ & 0.0063 \\
\hline
\end{tabular}
\caption{\label{tab:PyParameters} Permalloy material parameters used in micromagnetic simulations}
\end{table}

After defining the sample geometry, the system was given an initial magnetization pointing along the (0,0,1) direction out of the plane of the film and allowed to relax to the minimum energy state in the applied bias field, which ranged from 1 to 30 mT in our simulations.

To excite the system, we applied a Gaussian pulse with a 20 ps full width half maximum and a 0.5 mT amplitude. The system was then allowed to freely evolve in time for 20 ns and average magnetization was sampled every 5 ps. Finally, we used a discrete fast Fourier transform to analyze the data in the frequency domain. Using the above sampling rates and simulation lengths, we obtain a resolution of 50 MHz.

\begin{figure}
\includegraphics{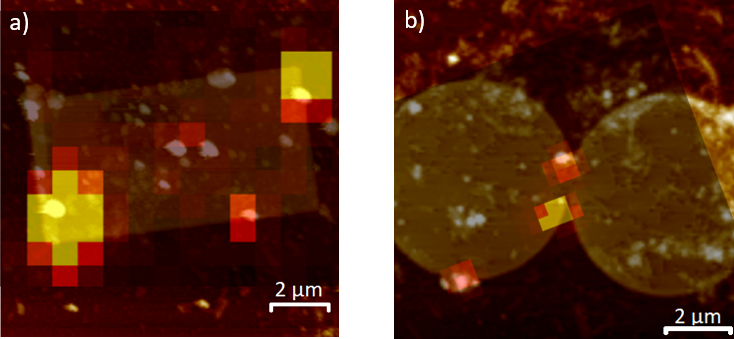}
\caption{\label{fig:AFMImages} AFM images of the measured permalloy features with superimposed scanning confocal fluorescence images showing where NV-containing nanodiamonds are located. (a) A 10 \textmu m x 5 \textmu m x 10 nm permalloy rectangle with nanodiamonds located near the edges and in the middle. (b) Two 6 \textmu m diameter, 10 nm thick permalloy circles connected at the edges. Nanodiamonds are located at the periphery of the left disk and near the junction between the disks. }
\end{figure}


\begin{figure}
\includegraphics{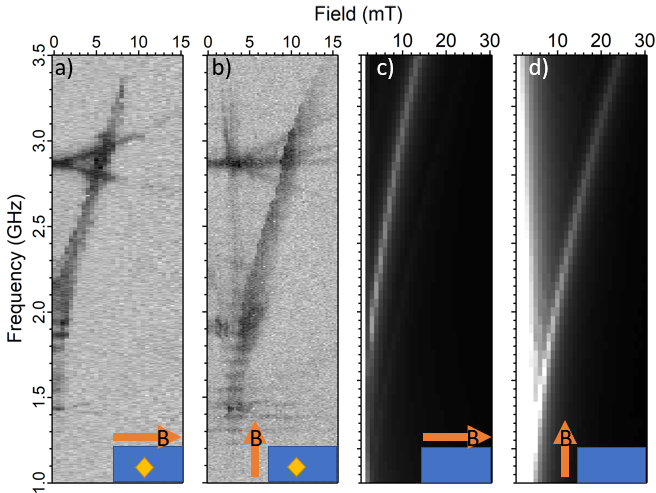}
\caption{\label{fig:RectBulk} Measurement of ODFMR using a nanodiamond in the bulk of a rectangular Py island with the applied magnetic field oriented along (a) the longitudinal easy axis and (b) along the transverse hard axis. The corresponding micromagnetic simulations of the FMR with the applied magnetic field oriented along (c) the longitudinal easy axis and (d) along the transverse hard axis.} 
\end{figure}

\section{\label{sec:level5}Results}
To begin, we placed multiple nanodiamonds on the 10 \textmu m x 5 \textmu m x 10 nm rectangle in Fig. \ref{fig:AFMImages} (a) to confirm that we could replicate previous macroscale FMR measurements carried out on arrays of such rectangular islands. These studies showed the presence of an easy-axis and hard-axis resonance \cite{skorohodov2017,zhang2016} which can be modelled using a geometry-dependent form of the Kittel equation:

\begin{equation*}
    f = \frac{\gamma}{2\pi}\sqrt{(H_\text{app}-(N_x - N_y)4\pi\mu_0 M_s)(H_\text{app}-(N_x - N_z)4\pi\mu_0M_s)}.
    \tag{2}
    \label{eqn:Kittel1}
\end{equation*}

Here, $\gamma$ is the gyromagnetic ratio of the material, $H_\text{app}$ is the applied field, $M_{s}$ is the saturation magnetization of the material, and $N_{i}$ are the three geometric demagnetization factors which sum to 1.\cite{Kittel1948} We use $N_x$ and $N_y$ as in-plane demagnetization factors and $N_z$ as the out-of-plane demagnetization factor. We can further simplify these three geometric parameters and the saturation magnetization into two quantities that represent the anisotropy fields of the features - $B_\parallel$ and $B_\perp$. This yields:

\begin{equation*}
    f = \frac{\gamma}{2\pi}\sqrt{(H_\text{app}+B_{\parallel})(H_\text{app}+B_{\perp})}.
    \tag{3}
    \label{eqn:Kittel2}
\end{equation*}

Because of Py's low intrinsic anisotropy, we can also assume that these anisotropy fields solely result from the shape anisotropy of our features. In the case of thin, rectangular features like the one measured, $N_x$ and $N_y$ in eqn. \ref{eqn:Kittel1} will be small and unequal, while $N_z$ will still be close to 1, its value in an infinite thin film. This results in $B_\perp$ being close to the $4\pi\mu_0M_s$ and in $B_\parallel$ having a comparatively small magnitude. Furthermore, when rotated 90$^{\circ}$ in plane, $N_x$ and $N_y$ swap positions in eqn. \ref{eqn:Kittel1}, leading to $B_\parallel$ retaining its magnitude while switching signs in \ref{eqn:Kittel2}. This results in a positive $B_\parallel$ along the easy axis, but negative $B_\parallel$ along the hard axis, leading to a divergence as $H_{app}$ approaches it.

Figure \ref{fig:RectBulk} (a) and (b) show the measurements of ODFMR obtained using a dim nanodiamond located near the center of the rectangle. When a magnetic field is applied along the easy axis (long edge of the rectangle), we detect a single mode that increases approximately linearly starting at 2 GHz (Fig. \ref{fig:RectBulk} (a)), as expected from eqn. \ref{eqn:Kittel2}. Fitting eqn. \ref{eqn:Kittel2} to this result, we find the in plane anisotropy field $B_\parallel$ of this feature to be 3.3 mT, and the out of plane anisotropy field $B_\perp$ to be 1.04 T, close to the accepted 1 T $M_s$ of Py as expected. When a magnetic field is applied along the hard axis (short edge of the rectangle), we see a V-shaped dispersion, which reaches a minimum at approximately 3 mT, near the divergence point expected from our previously measured $B_\parallel$ (Fig. \ref{fig:RectBulk} (b)). The micromagnetic simulations shown in Fig. \ref{fig:RectBulk} (c) and (d) largely match these results, though the frequencies are higher, most likely a result of our applied microwave power, which can result in NV contrast above the FMR frequency \cite{Mccullian2020}.

\begin{figure}
\includegraphics{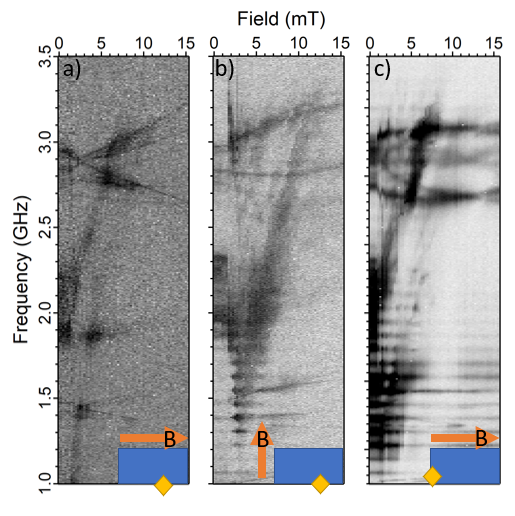}
\caption{\label{fig:RectEdge} ODFMR measurements using a nanodiamond located on the long edge of a rectangular island, with field oriented along (a) the hard axis (long edge), (b)  along the easy axis (short edge). (c) ODFMR measurements using a nanodiamond located on the short edge of a rectangular island with field oriented along the hard axis (long edge). We attribute the anomalous high field behavior of the NV center resonance lines to fringe field effects that occur at the edge of the patterned feature.}
\end{figure}

However, along the edges of the rectangle, we detected new features in addition to the ones seen by the nanodiamond located in the middle (or bulk) of the ferromagnetic rectangle  (Fig. \ref{fig:RectEdge}). For a nanodiamond located on the long edge, when the magnetic field is applied along the easy axis (long edge of the rectangle), we observe an incomplete switching with a faint signal from the V-shaped resonance  (Fig. \ref{fig:RectEdge} (a). We believe this to be the result of a small magnetic field misalignment. When the field is applied along the hard axis (Fig. \ref{fig:RectEdge} (b)), we see a faint easy-axis signal as well as faint traces of additional signals between the easy and hard axis FMR signals. For a nanodiamond placed on the short edge of the rectangle and with the field applied along the easy axis Fig. \ref{fig:RectEdge} (c)), we observe a strong easy axis signal and a faint hard axis signal, similar to Fig. \ref{fig:RectEdge} (a). However, we can also see faint traces of additional signals above the easy-axis resonance. We speculate that these additional signals in Fig. \ref{fig:RectEdge}(b) and (c) are higher order magnon modes.

\begin{figure}
\includegraphics{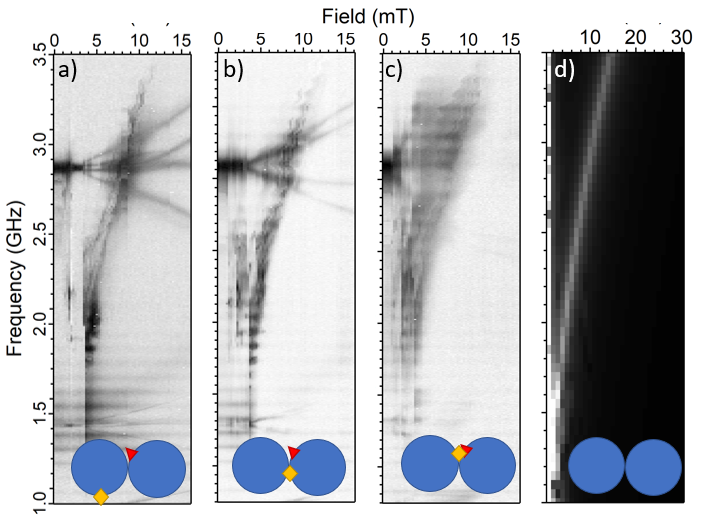}
\caption{\label{fig:CircleImage} ODFMR measurements using a nanodiamond located at different sites along the edge of a circular double disk feature where the disks meet to form a constriction. Measurements are made at three distinct sites: (a) Far away from the constriction;(b) Near the constriction; (c) On the opposite end of the constriction, close to a misfabricated edge. (d) Micromagnetic simulations of FMR in the double circle feature.}
\end{figure}

We now discuss measurements on a feature composed of two connected 6 \textmu m diameter, 10 nm thick circular Py disks (Fig. \ref{fig:AFMImages} (b)). These features had a slight fabrication error in one of the circles near the constriction where they meet, leading to a sharp edge slightly protruding, which could cause a local distortion of the ferromagnetic resonance. We positioned and measured nanodiamonds at three different sites - one on the edge approximately 4 \textmu m from the constriction, one on the pristine side of the constriction approximately 2 \textmu m from the fabrication error, and one on the misfabricated side of the constriction, approximately 750 nm away from the error. Far away from the constriction, the data shows streaking at approximately 2 mT, and a signal that begins suddenly at approximately 4 mT (Fig. \ref{fig:CircleImage} (a)). Near the constriction (Fig. \ref{fig:CircleImage} (b)), we see additional noise broadband noise at lower fields (between 2 mT - 4 mT), but the signal largely matches that in Fig. \ref{fig:CircleImage} (a). On the opposite end of the constriction (Fig. \ref{fig:CircleImage} (c)), close to the misfabricated edge, the primary resonance detected at higher fields matches the measurements at the other two sites, but an additional high frequency resonance emerges, leaving our detection range between 5 and 10 mT. At this site, the signal up to 5 mT appears to be broadband noise, with greater contrast as the field increases. The line width also increases dramatically, from 156 MHz at 10 mT at the opposite, pristine side of the constriction to 278 MHz at 10 mT. Micromagnetic simulations of FMR in the double circle feature (\ref{fig:CircleImage} (d)) match our experimental results well for fields stronger than 4 mT. We speculate that the disagreement between measurements and simulations at lower fields is caused by differences in the magnetic texture of the material, as the applied field will not saturate the ferromagnet and our simulation setup procedure does not perfectly replicate the history of the features.

\section{\label{sec:level6}Summary}
We have demonstrated a straightforward method to deterministically place NV-containing nanodiamonds at desired locations on lithographically patterned ferromagnetic thin films. Using this placement, we performed local ODFMR measurements on single mesoscopic Py islands, revealing position-dependent variations in spin dynamical behavior that cannot be detected using conventional FMR measurements of ensembles of patterned islands. After confirming that this technique worked on a simple rectangular island, we then applied it to a more complex feature composed of two circles with a subtle fabrication error near the point of closest approach. In the vicinity of this fabrication error, we measured both a larger line width and an additional signal that did not appear in the control measurement away from the error or in our micromagnetic simulations of pristine patterns. This finding shows that local ODMFR measurements using targeted placement of nanodiamonds can provide information about the influence of defects on the spin dynamical behavior of patterned ferromagnets.   

Moving forward, we see this technique being used as a more general method of measuring localized magnetization dynamics in various patterned ferromagnetic structures. For example, the technique could provide new insights into the properties of spin wave edge modes previously detected using scanning FMRFM \cite{Guo_PhysRevLett.110.017601}. It could also be used for probing the local magnetization dynamics of artificial spin ice arrays \cite{Bhat_PhysRevLett.125.117208,Kempinger2021}. Finally, because we have achieved sub-100 nm placement accuracy, we believe this technique could be used to more readily explore the dynamics of single trapped domain walls, skyrmions, or other nanoscale magnetic textures in samples where relying on either stochastic assembly or templated assembly may not be feasible.

\begin{acknowledgments}
The authors thank Eric Kamp for initiating the NV center project in our research group and David Awschalom for useful discussions. We are grateful to Michael Labella for his sample fabrication advice. We acknowledge support from the University of Chicago and the U.S. Department of Energy Office of Science National Quantum Information Science Research Centers (Q-NEXT).
\end{acknowledgments}

%

\end{document}